# Wavefront sensing with a Gradient Phase Filter

François Hénault[1], Yan Feng[2], Jean-Jacques Correia[1], Laura Schreiber[1,3], Alain Spang[4]

[1] Institut de Planétologie et d'Astrophysique de Grenoble, Université Grenoble-Alpes, CNRS, 38041 Grenoble – France
[2] Laboratory of Biology, Engineering and Imaging for Ophthalmology (BiiO), Faculty of Medicine, Health Innovation campus, Jean Monnet University, Saint-Etienne – France
[3] INAF - Osservatorio di Astrosica e Scienza dello Spazio di Bologna, Via Gobetti 93/3, 40129 Bologna – Italy
[4] Université Côte d'Azur, Observatoire de la Côte d'Azur, CNRS, Laboratoire Lagrange – France

E-mail: henaultfbm@gmail.com

## ABSTRACT

Wavefront sensors have now become core components in the fields of metrology of optical systems, biomedical optics, and adaptive optics systems for astronomy. However, none of the designs used or proposed so far achieve simultaneously a high spatial resolution at the pupil of the tested optics and absolute measurement accuracy comparable to those of modern laser-interferometers. This paper presents an improved wavefront sensor concept that reaches both previous goals. This device named Crossed-sine phase sensor (CSPS) is based on a fully transparent gradient phase filter (GPF) placed at an intermediate location between the virtual pupil and image planes of the tested optics. The theoretical principle of the sensor is described in Fourier optics formalism. Numerical simulations confirm that a measurement accuracy of $\lambda/100$ RMS is achievable. The CSPS also offers the advantages of being quasi-achromatic and working on spatially or spectrally extended, natural or artificial light sources.

## 1  INTRODUCTION

Throughout the years, Adaptive optics (AO) systems have become essential for various applications such as the metrology of optical components or systems [1-3], biomedical optics and ophthalmology for cornea or retinal diagnostics [4-5], astronomical observations [6], and characterization of laser beams [7]. The core element of these systems is undoubtedly the Wavefront sensor (WFS) measuring an incoming distorted wavefront in real time and sending this information to a deformable mirror compensating for the input Wavefront errors (WFEs). Nowadays the most popular WFS are the Shack-Hartmann (SH) based on a micro-lens array placed at the pupil of the tested optical system [8] and the Pyramid sensor (PS) that combines four pupil images seen through a faceted prism located at the focus of the system [9]. Less known are the Optical differentiation sensor (ODS) making use of a Gradient density filter (GDF) located in the image plane of the tested optical system [10] and the Multi-wave lateral shearing interferometer [11] based on the implementation of a chessboard phase grating [12]. All these concepts present some drawbacks. For example, the SH WFS is limited in terms of WFE measurement accuracy and spatial resolution due to the pitch of its micro-lenses array. Conversely, the PS can achieve a high spatial resolution that is only limited by the pixels number of the detector array recording the pupil images. Its main disadvantage is the need to modulate the acquired signals, thus moving the pyramid prism continuously and leading to technical difficulties. Because the ODS makes use of a Gradient density filter (GDF, it is not well suited to astronomical observations [13-15]. Recently, its original concept was improved by replacing the usual on-axis light source with four off-axis ones whose images are acquired simultaneously, thus giving a multiplex advantage. This concept allows reaching measurement accuracy comparable to those of Fizeau laser-interferometers [16-17]. However this design named Crossed-sine sensor (CSS) still makes use of a GDF, thus implying throughput losses.

In this paper the GDF is replaced with a Gradient phase filter (GPF) ensuring maximal luminosity of the acquired pupil images. Referring to the CSS, this new concept was renamed as the Crossed-sine phase sensor (CSPS). It is insensitive to disturbances generated by the environment such as micro vibrations and atmospheric turbulence, only requires standard optical components, and is applicable to natural or artificial light sources, being spatially and spectrally extended or not. The paper presents the optical concept and design of the CSPS. The experimental validation of this concept will be the subject of a future paper. It is firstly described in section 2, including the GPF and two possible optical schemes. Section 3 explains its theoretical principle in the framework of Fourier optics as well as the derived data processing algorithm. The achievable measurement accuracy is evaluated with the support of numerical simulations presented in section 4. The general properties of the CSPS are summarized in the concluding section 5.

## 2 THE CROSSED-SINE PHASE SENSOR: CONCEPT AND DESIGN

The CSPS concept makes use of a gradient phase filter (GPF) to modulate the complex electric field sliced by a lenslet behind an image plane. The CSPS takes its inspiration from the reverse Hartmann test [2, 18] and from optical differentiation sensors [10]. These sensors make use of a real or virtual diffraction grating located between the exit pupil of the tested optics and its real image plane. This diffraction grating can either be a classical one or a volume phase holographic grating (VPHG) appended to it. This grating generates a series of ghost images of the exit pupil in which the input wavefront is coded into varying intensities. The case of the phase grating described in this paper, however, involves infinite series of Bessel functions, thus adding not negligible complexity to the previous concepts. To support the reader in concept understanding, we report in Section 3.1.2 a simplified graphical interpretation. Nevertheless, the data reduction process presented in section 3.2 allows solving this difficulty.

### 2.1 Design description

The CSPS design essentially comprises two elements: an optical measurement head and a calculation unit (Figure 1):
- The optical measurement head simultaneously acquires several greyscale images of the pupil plane. It is composed of the GPF and of an optical re-imaging system that includes a lenslet array.
- The calculation unit is used for digital reconstruction of the WFE from the acquired images.

These two elements are described in detail in the next sections.

#### 2.1.1 Coordinate systems and scientific notations

In this section we describe the employed Cartesian coordinate depicted in Figure 2.
- The OXYZ reference frame is attached to the exit pupil of the tested optical system. The point O is located at the pupil centre and OZ is the optical axis of the system. Points P in the pupil plane OXY are denoted by their Cartesian coordinates $(x,y)$ and the WFE to be measured is noted $\Delta(x,y)$.

- The O'X'Y'Z reference frame is attached to the focal plane of the tested optics. Point O' is the nominal focus located at a distance $f' = OO'$ from the pupil, with $f$ the focal length of the system. Points M' located in the O'X'Y' plane are denoted by their Cartesian coordinates $(x',y')$.
- The O"X"Y"Z reference frame is attached to the plane of the detector array where the pupil images are formed. Points P" located in the O"X"Y" plane are denoted by their Cartesian coordinates $(x", y")$. Since they are strictly homothetic to the coordinates $(x,y)$, they will be replaced in the remainder of the texts.

Moreover, two additional coordinate systems are attached to the GPF and to its virtual image through the optics (see Figure 2):
- The $FX_FY_FZ$ reference frame at the location of the real, physical GPF with Cartesian coordinates $(x_F, y_F)$.
- The F'X'$_V$Y'$_V$Z reference frame at the location of its virtual image between the OXY and O'X'Y' planes, with Cartesian coordinates $(x'_V, y'_V)$.

## 2.2 Gradient Phase Filter (GPF)

The GPF constitutes the essential element of the CSPS and it is located inside the optical measurement head. Its transmission is expressed by a complex function defined mathematically as the product of two sinusoidal phase functions:

$$T_F(x_F, y_F) = \exp\left[2j\pi\varphi_0 \sin\left(\sqrt{2}\pi(x_F - y_F)/p\right)\sin\left(\sqrt{2}\pi(x_F + y_F)/p\right)\right] \quad (1)$$

that also writes as:

$$T_F(x_F, y_F) = \exp\left[j\pi\varphi_0 \left(\cos(2\sqrt{2}\pi y_F/p) - \cos(2\sqrt{2}\pi x_F/p)\right)\right] \quad (2)$$

where $j$ is the complex root of $-1$, $\lambda$ the operating wavelength, $\varphi_0$ the amplitude of the phase gradient, and $p$ the spatial period of the filter assumed to be identical in both directions $X_F$ and $Y_F$. It should be noted that this component belongs to the family of VPHGs, which are commonly used in the fields of spectroscopy and astronomy [19-20]. Thus they are inexpensive and available from many suppliers.

The function $T_F(x_F, y_F)$ is illustrated with false colors in Figure 3. It shows the locations of the observation points $F_i$ ($1 \le i \le 9$) at the GPF plane. They are denoted by the vectors **FF$_i$** and are defined as follows:

- They can be located along the $X_F$ and $Y_F$ axes as indicated by light blue dots in Figure 3 (i = 2, 4, 6 and 8),
- Their locations can be rotated by an angle of 45 degrees with respect to the $X_F$ and $Y_F$ axes as indicated by dark blue dots in Figure3 (i = 1, 3, 7 and 9),
- It must be noted that any combination of these points is possible, thus involving 4, 5, 8 or 9 different observation points. A central observation point noted $F_5$ coincides with point F and may be used for radiometric calibration purpose.

## 2.3 Optical schemes

The main functions of the tentative optical layouts described below are as follows:
- Creating a virtual image of the GPF located in front or behind the exit pupil of the tested optics,
- Ensuring the capture of different pupil images seen from the observation points L'$_i$ located in the image plane O'X'Y' ($1 \le i \le 9$),
- Preventing any pupil images overlap in the plane of the detector array.

Two different optical schemes denoted V1 and V2 are depicted in Figure 2. They show the distances between the main optical components, including the lenslet array and the real GPF and its virtual image. The nomenclature of these parameters is given in Table 1. They are of two different kinds: basic parameters of the experiment, such as the wavelength of the incoming radiation and the numerical aperture of the tested optics, and free parameters to be optimized in order to improving the WFE measurement accuracy. These are marked with asterisks in Table 1. They include the spatial period $p$ of the GPF, the amplitude $\varphi_0$ expressed as a $\pi$-fraction and the distance $z'$ from its virtual image to the image plane O'X'Y'.

Each of these optical schemes has its own advantages and drawbacks. The V1 scheme presents the best radiometric transmission. In return, it uses an unconventional lenslet array where the optical axes of the mini lenses are shifted laterally with respect to their contours. The V2 scheme only makes use of conventional optical components (the lenslet array and a telecentric lens). Conversely, its radiometric transmission is slightly lower.

It should be noted that two other options were envisaged: the first one consists in inverting the locations of the lenslet array and the telecentric lens, and the second in replacing that lens with a second lenslet array. None of them exhibited better performance or simpler optical design, thus they are not presented here. We finally selected the V2 optical scheme as the reference, which is studied in the remainder of the text.

Table 1: Nomenclature and numerical values of the parameters used in numerical simulations.

|  | Symbol | Value | Unit |
|---|---|---|---|
| Wavelength of incoming electromagnetic radiation | $\lambda$ | 0.5 | micron |
| Wavenumber of incoming electromagnetic radiation | $k = 2\pi/\lambda$ | 20000 | cm$^{-1}$ |
| GPF spatial period (*) | $p$ | 1 | mm |
| GPF phase amplitude (*) | $\varphi_0$ | $\pi$ | rad |
| Focal length of tested optical system | $f$ | 500 | mm |
| Aperture number of tested optical system | $N$ | 20 | - |
| Lenslet array dimensions | $l_M \times h_M$ | 1 x 1 | mm |
| Focal length of lenslet array (*) | $f_M$ | 28.5 | mm |
| Distance from image plane O'X'Y' to lenslet array (*) | $z_M$ | 1 | mm |
| Distance from lenslet array to GPF | $z_F$ | 31.5 | mm |
| Distance from image plane O'X'Y' to virtual GPF (*) | $z'$ | -200 | mm |
| Distance from GPF to second lens | $z_2$ | 23.3 | mm |
| Distance from second lens to detector array | $z''$ | 37 | mm |

(*) Optimized parameters

## 3 THEORETICAL ANALYSIS

### 3.1 Fresnel diffraction analysis

Finding analytical expressions of the pupil images I''$_i$(x'',y'') formed on the detector array (with $1 \leq i \leq 9$) is not very difficult but long and cumbersome. They are obtained from a Fresnel diffraction analysis that was described in Ref. [18] and is summarized blow. It makes use of mathematical and condensed notations defined in Table 2.

Table 2: Nomenclature of mathematical symbols and their condensed notations.

| Symbol | Mathematical notation | Condensed notation |
|---|---|---|
| Pupil transmission map in the exit pupil plane OXY of the tested optics. It is defined to be a pillbox function equal to unity inside a disk of diameter $D$ and to zero outside of it | $B_D(x, y)$ | $B_D$ |
| Wavefront error (WFE) to be measured | $\Delta(x'', y'')$ | $\Delta$ |
| WFE slopes along X-axis | $\partial \Delta(x'', y'')/\partial x''$ | $\Delta'_X$ |
| WFE slopes along Y-axis | $\partial \Delta(x'', y'')/\partial y''$ | $\Delta'_Y$ |
| Bessel functions of the first kind at the m$^{th}$ order | $J_m(\pi \varphi_0)$ | $J_m$ |
| Phase-shifts of off-axis images from the O'X'Y plane | $\phi = \pm \pi/2$ | – |

*3.1.1 On-axis observation point*

In Fresnel diffraction theory, the complex amplitude diffracted in the plane of the virtual GPF can be expressed as [21]:

$$A'_V(x'_V, y'_V) = -\frac{i}{\lambda d''} \exp[ikd''] \exp[ik(x_V'^2 + y_V'^2)/2d''] \iint_{x,y} B_D \exp[ik\Delta] \exp[-ik(xx'_V + yy'_V)/d''] \exp[ik(x^2 + y^2)/2d''] dxdy \quad (3)$$

where $d'' = f + z'$ is the Fresnel diffraction distance from the pupil to the virtual filter plane F'X'$_V$Y'$_V$. Similarly, a reversed Fresnel diffraction operation allows expressing the complex amplitude $A''(x'', y'')$ formed in the plane of the detector array, which is optically conjugated with the exit pupil of the tested optics:

$$A''(x'', y'') = \frac{1}{\lambda^2 d''^2} \iint_{x,y} B_D \exp[ik(\Delta + (xx'_V + yy'_V)/d'' + (x^2 + y^2)/2d'')] \\ \times \left( \iint_{x'_V, y'_V} T'_V(x'_V, y'_V) \exp[ik(x'_V(x'' - x) + y'_V(y'' - y))/d''] dx'_V dy'_V \right) dxdy \quad (4)$$

where the multiplying phase terms are omitted because they will disappear when computing the image intensity, and $T'_V(x'_V, y'_V)$ is the complex transmission of the virtual GPF derived from Equation 2:

$$T'_V(x'_V, y'_V) = \exp\left[j\pi\varphi_0 \left(\cos(2\sqrt{2}\pi y'_V/p'_V) - \cos(2\sqrt{2}\pi x'_V/p'_V)\right)\right] \quad (5)$$

Referring to the V2 optical scheme in Figure 2, the spatial period of the virtual GPF is equal to $p'_V = p(f - z')/(f + z_F)$. This expression of $T'_V(x'_V, y'_V)$ can be developed into a series of Bessel functions by use of the Jacobi-Anger expansion, stating that for any $z$ and $\theta$ parameters [22]:

$$\exp(jz\cos\theta) = J_0(z) + 2\sum_{m=1}^{+\infty} (-1)^m J_m(z) \cos(m\theta), \quad (6)$$

where $J_m(z)$ are Bessel functions of the first kind at the m$^{th}$ order. Using the condensed notations in Table 2, Equation 6 can be rewritten as:

$$T'_V(x'_V, y'_V) = J_0^2 + 2J_0 \left( \sum_{m=1}^{+\infty}(-1)^m J_m \cos(mu) + \sum_{n=1}^{+\infty}(-1)^n J_n \cos(nv) \right)$$
$$+ 4 \sum_{\substack{m=1 \\ m>n}}^{+\infty} \sum_{\substack{n=1 \\ n>m}}^{+\infty} (-1)^{m+n} J_m J_n \cos(mu)\cos(nv) \quad , \tag{7a}$$

With:

$$u = 2\sqrt{2}\pi x'_V / p'_V$$
$$v = 2\sqrt{2}\pi y'_V / p'_V \quad . \tag{7b}$$

Inserting this last expression of $T'_V(x'_V, y'_V)$ into Equation 4, then reducing the arguments of the complex exponentials and reordering the summation operators leads to the mathematical development that is described into the Appendix 1. Then the expression of the complex amplitude distribution in the O"X"Y" $A''(x'', y'')$ can be written as:

$$A''(x'', y'') = \left\{ J_0^2 + 2J_0 \sum_{m=1}^{+\infty}(-1)^m J_m \cos\left[k\left(mp''\Delta'_X + \frac{mp''}{d''}x''\right)\right] \right.$$
$$+ 2J_0 \sum_{n=1}^{+\infty}(-1)^n J_n \cos\left[k(np''\Delta'_Y) + \frac{np''}{d''}y''\right]$$
$$+ 2 \sum_{\substack{m=1 \\ m>n}}^{+\infty} \sum_{\substack{n=1 \\ n>m}}^{+\infty} (-1)^{m+n} J_m J_n \cos\left[ik\left(mp''\Delta'_X + np''\Delta'_Y + p''\frac{mx'' + ny''}{d''}\right)\right]$$
$$+ 2 \sum_{\substack{m=1 \\ m>n}}^{+\infty} \sum_{\substack{n=1 \\ n>m}}^{+\infty} (-1)^{m+n} J_m J_n \cos\left[ik\left(mp''\Delta'_X - np''\Delta'_Y + p''\frac{mx'' - ny''}{d''}\right)\right] \tag{8}$$

It is now possible to calculate the image intensity of the central image $I''(x'', y'')$ by taking the square modulus of the complex amplitude $A''(x'', y'')$. This mathematical development is not extremely difficult but very long and cumbersome. It will be found that $I''(x'', y'')$ is composed of quasi-infinite series of cosine terms proportional to fourth-order coefficients with respect to $J_0$, $J_m$ and $J_n$, i.e. $J_0^4, J_0^3 J_m, J_0^3 J_n, J_0^2 J_m J_n, J_0 J_m^2 J_n, J_0 J_m J_n^2$ and $J_m^2 J_n^2$. Instead of giving a complete expression of $I''(x'', y'')$ here, we should firstly simplify it with the help of a graphical interpretation that is described below.

### 3.1.2 Graphical interpretation

Figure 4 shows different histograms of the Bessel function products for the cases when $\varphi_0 = \pi/2$ (Figure 4-a), $\varphi_0 = \pi$ (Figure 4-b) and $\varphi_0 = 3\pi/2$ (Figure 4-c). Figure 4-d demonstrates that the most powerful component is observed in the range $\varphi_0 \in [0.5\pi - 0.6\pi]$, and that high order terms become negligible when $\varphi_0 > \pi$. A Taylor expansion of Equation 8 at the fourth order with respect to $J_1$ only leads to the following analytical approximation:

The complex amplitude $A''(x'', y'')$ formed on-axis:

$$A''(x'', y'') = 2J_1^2 \cos\left(\frac{x''+y''}{d''}\right)\cos\left[kp''\left(\Delta'_{X'} + \Delta'_Y + \frac{x''+y''+2p''}{d''}\right)\right] + 2J_1^2 \cos\left(\frac{x''-y''}{d''}\right)\cos\left[kp''\left(\Delta'_{X'} - \Delta'_Y + \frac{x''-y''}{d''}\right)\right], \quad (9)$$

and the resulting on-axis intensity writes as:

$$I''(x'', y'') = 2A\cos^2\left[kp''\left(\Delta'_{X'} + \Delta'_Y + \frac{x''+y''+2p''}{d''}\right)\right] + 2B\cos^2\left[kp''\left(\Delta'_{X'} - \Delta'_Y + \frac{x''-y''}{d''}\right)\right] + 2C\cos\left[kp''\left(\Delta'_{X'} + \Delta'_Y + \frac{x''+y''+2p''}{d''}\right)\right]\cos\left[kp''\left(\Delta'_{X'} - \Delta'_Y + \frac{x''-y''}{d''}\right)\right] \quad (10)$$

with: $A = 2J_1^4 \cos^2\left(\frac{x''+y''}{d''}\right)$, $B = 2J_1^4 \cos^2\left(\frac{x''-y''}{d''}\right)$ and $C = 4J_1^4 \cos\left(\frac{x''+y''}{d''}\right)\cos\left(\frac{x''-y''}{d''}\right)$

*3.1.3 Off-axis observation points*

The coordinates of the observation points $L_i$ ($1 \leq i \leq 9$), into the image plane O'X'Y' are defined in Figure 2 by the vectors $\mathbf{O'L'_i}$ whose coordinates are defined as:

$$\mathbf{O'L'_i} = \frac{f}{G}\begin{pmatrix}\pm\pi/2 + 2m\pi \\ \pm\pi/2 + 2n\pi\end{pmatrix} \quad \text{for points located along the } X_F \text{ and } Y_F \text{ axes } (i = 2, 4, 6, 8)$$

$$\mathbf{O'L'_i} = \frac{f}{G}\begin{pmatrix}\pm\pi/2 + 2m\pi + 2n\pi \\ \pm\pi/2 + 2m\pi + 2n\pi\end{pmatrix} \quad \text{for points located at 45 degrees from these axes } (i = 1, 3, 7, 9) \quad (11)$$

where *m* and *n* are non-zero integers and *G* is defined as the gain factor of the CSPS. Given the optical layout V2 described in Figure 2-V2 and the analytic notations in Table 1 it is equal to:

$$G = 2\pi\sqrt{2}\, z_F \frac{f+z'}{p\, z'}. \quad (12)$$

It may be noted that when projected into the virtual GPF plane F'X'$_V$Y'$_V$ the points $L_i$ correspond to the offset red points $F_i$ linked with a vectorial relation:

$$\mathbf{FF_i} = \frac{f+z_F}{f}\mathbf{O'L'_i}. \quad (13)$$

From the off-axis observation points the WFE to be measured includes additional tilt terms writing as:

$$W(x'', y'') = \Delta(x'', y'') + \frac{\mathbf{O'L'_i} \cdot \mathbf{s'}}{f} \quad (14)$$

with $1 \leq i \leq 9$, $\mathbf{s'}$ is a vector of coordinates $(x', y')$, and the coordinates of vectors $\mathbf{O'L'_i}$ are those defined by Equations 11. They are defined so to generate phase-shifts equal to $\phi = \pm\pi/2$ with respect to the central image, i.e.

    $\phi = \pm\pi/2$     along X–axis for images numbers 4 and 6,
    $\phi = \pm\pi/2$     along Y–axis for images numbers 2 and 8,
    $\phi = \pm\pi/2$     along both X and Y axes for images numbers 1, 3, 7 and 9.

Therefore some cosine terms in Equation 10 will turn into sine functions and the expressions of the intensity distributions observed at the pupil plane from the off-axis observation points are given into Appendix 2.

In principle, combining the previous expressions of $I''_{ki}(x'', y'')$ with $1 \leq i \leq 9$ allows determining all their sine terms, and then their arguments that are proportional to the searched WFE slopes $\Delta'_X$ and $\Delta'_Y$. Practically speaking however, such analytical expressions are approximate, thus an alternative way to extract more precise data is described in the next subsection.

### 3.2 Data reduction process

Here we make use of a double Fourier transformation algorithm that was developed previously in the framework of reverse Hartmann wavefront sensing [18]. It consists in projecting the complex amplitude from the exit pupil plane of the tested optics to the virtual GPF via a Fresnel transform, multiplying it with the GPF transmission, and returning into the pupil plane via a reverse Fresnel transform to get the final images. The data reduction process is described below and illustrated in Figure 5.

1) Start with one of the acquired images $I''_i(x'', y'')$ with $1 \leq I \leq 9$,

2) Calculate the Fourier transformation of $I''_i(x'', y'')$,

3) As illustrated in Figure 5, isolate the first satellite peak of this Fourier transform, which is located along the horizontal axis U at the spatial frequency $\delta u = G/2\pi d''$,

4) Recentre this peak on the origin of the UV plane,

5) Calculate the inverse Fourier transform of the re-centered peak. The result is a complex function spanning into the O"X"Y" plane,

6) Extract the phase of this complex function, which is equal to $-j\, G\, \Delta'_X /2$,

7) Repeat step 3 by isolating the first satellite peak along the vertical axis V at the spatial frequency $\delta v = G/2\pi d''$,

8) Repeat steps 4 to 6. The phase of the complex function is now equal to $-j\, G\, \Delta'_Y /2$,

9) Repeat all operations 1 to 8 for all the acquired images $I''_i(x'', y'')$ with $1 \leq i \leq 9$.

Let us note that some preliminary processing of the acquired images $I''_i(x'', y'')$ is needed, such as:

  - Correction of the background and spatial non-uniformity of the detector matrix (dark noise and flat field),

  - Separation of the sub-images $I''_i(x'', y'')$ that are stored into different arrays,

  - Re-centring these sub-images on point O" (registration).

Finally, the WFE $\Delta(x'', y'')$ can be reconstructed from its slopes $\Delta'_X(x'', y'')$ and $\Delta'_Y(x'', y'')$ by use of some classical algorithms such as Zernike modal fitting [23] or iterative Fourier transforms [24]. Here the second WFE reconstruction procedure is selected in view of its shorter computation time. Numerical simulations were carried out following that procedure, and the results are summarized in the next section.

# 4 NUMERICAL SIMULATIONS

The numerical simulations presented here make use of a complex amplitude propagation model of Fresnel diffraction that was described in Ref. [19]. They were made in the absence of detector which will be included in a future work. They are limited to the four "diagonal" images $I_1''(x'',y'')$, $I_3''(x'',y'')$, $I_7''(x'',y'')$ and $I_9''(x'',y'')$ that are depicted in Figure 6. It has been checked that similar results are obtained with the "horizontal" and "vertical" images $I_2''(x'',y'')$, $I_4''(x'',y'')$, $I_6''(x'',y'')$ and $I_8''(x'',y'')$. The simulation results are summarized in Table 3 and Table 4 for the cases of low and mid-order Zernike polynomials, respectively. The tables show the Peak-to-Valley (PTV) values and the RMS standard deviations of the achieved measurement accuracy. These results are illustrated in Figure 7 and Figure 8 which show false color views of the WFE slopes along the X" and Y" axes, the reconstructed WFE, and their differences with respect to the reference case. They were built from randomly defined coefficients of the first 16 Zernike polynomials (low-order case) and 48 Zernike polynomials (mid-order case). Their PTV and RMS values are equal to 0.114λ PTV and to 0.016λ RMS, and to 0.438λ PTV and 0.054λ RMS respectively. This type of wavefront errors is characteristic of optical aberrations or of mechanical deformation modes of mirrors or lenses. It also stands for differential errors between the science and metrology channels of an AO system.

The numerical simulations with low-order Zernike polynomials show that the absolute measurement error is about λ/120 RMS, which corresponds to relative errors of 5% RMS. A slightly worse performance is detected in measuring higher order Zernike polynomials. These figures are comparable to the precision achieved by laser-interferometers. One may also note in Figure 7 that the largest errors are located near to the pupil rim, meaning that better values are achievable over slightly reduced measurement areas (typically by 5% of the pupil diameter).

Table 3: Numerical results of the CSPS simulations with low-order Zernike polynomials ≤ 16.

| Error type | Reference | Measured | Difference | Relative Difference (%) | |
|---|---|---|---|---|---|
| X-slopes (μrad) | 130 | 137 | 39 | 30 | PV |
|  | 25 | 25 | 4 | 17 | RMS |
| Y-slopes (μrad) | 151 | 157 | 51 | 34 | PV |
|  | 30 | 30 | 7 | 23 | RMS |
| Wavefront Error (waves) | 0.787 | 0.785 | 0.057 | 7 | PV |
|  | 0.162 | 0.162 | 0.008 | 5 | RMS |

Table 4: Numerical results of the CSPS with mid-order Zernike polynomials ≤ 48.

| Error type | Reference | Measured | Difference | Relative Difference (%) | |
|---|---|---|---|---|---|
| X-slopes (µrad) | 190 | 180 | 75 | 30 | PV |
| | 39 | 39 | 11 | 28 | RMS |
| Y-slopes (µrad) | 273 | 233 | 101 | 37 | PV |
| | 48 | 49 | 14 | 29 | RMS |
| Wavefront Error (waves) | 1,013 | 0,979 | 0,219 | 22 | PV |
| | 0,216 | 0,214 | 0,027 | 13 | RMS |

## 5 CONCLUSION AND FUTURE WORK

This paper presents the design of a crossed-sine phase wavefront sensor which is based on a fully transparent gradient phase filter located between the virtual pupil and image planes of the tested optical system. It allows measuring and controlling the optical wavefronts on spatially or spectrally extended objects, either of the natural or artificial types. Its main features are as follows:

- It makes it possible to achieve measurement accuracy comparable to that of laser-interferometers, i.e. one hundredth of the measurement wavelength typically,
- It provides access to very high spatial resolution equivalent to several million pixels on the surface of the tested optical system,
- It allows measurements to be carried out in a short time period (typically < 0.01 second) to overcome disturbances generated by the environment in the field of ophthalmology or astronomical applications,
- Finally, it comes under the form of a small measurement head (maximum length ≤ 50 mm) that can easily be integrated into optical systems.

The theoretical principle of the sensor was described in Fourier optics formalism. Numerical simulations confirmed that a measurement accuracy of $\lambda/100$ RMS is achievable with spatial resolutions about 1000 x 1000 or more. This makes the crossed-sine phase sensor a good candidate for optical metrology, ophthalmology or astronomical applications in the future. In this prospect the development of a demonstrating prototype has been started.

Further improvement of both the CSPS system and GPF is highly expected in the future, i.e. a ray-tracing model of the measurement head, characterisation of the CSPS system (such as dynamic range, linearity and so on) and more precise design of GPF. At a later stage, we will make a CSPS prototype and compare its performance with classical SH, pyramid and multi-wave lateral shearing interferometer WFS designs.

# APPENDIX 1. EXPRESSING THE COMPLEX AMPLITUDE $A''(x'',y'')$ ON-AXIS

Inserting the expression of $T'_V(x'_V, y'_V)$ in Equation 2 of the main text into relation 4, them reducing the arguments of the complex exponentials, and reordering the summation operators leads to:

$$\begin{aligned} A''(x'',y'') = \frac{1}{\lambda^2 d''^2} \Bigg\{ & J_0^2 \frac{p_V''^2}{\pi^2} \iint_{x,y} B_D \exp\left[ik\left(\Delta + (x^2+y^2)/2d''\right)\right] \left(\iint_{u,v} \exp\left[ik\, p'_V(ux''+vy'')/2\sqrt{2}\pi d''\right] du\, dv\right) dx\, dy \\ & + J_0 \frac{p_V''^2}{4\pi^2} \sum_{m=1}^{+\infty} (-1)^m J_m \iint_{x,y} B_D \exp\left[ik\left(\Delta + (x^2+y^2)/2d''\right)\right] \left(\iint_{u,v} \exp\left[ik\, p'_V(ux''+vy'')/2\sqrt{2}\pi d''\right] \cos(mu)\, du\, dv\right) dx\, dy \\ & + J_0 \frac{p_V''^2}{4\pi^2} \sum_{n=1}^{+\infty} (-1)^n J_n \iint_{x,y} B_D \exp\left[ik\left(\Delta + (x^2+y^2)/2d''\right)\right] \left(\iint_{u,v} \exp\left[ik\, p'_V(ux''+vy'')/2\sqrt{2}\pi d''\right] \cos(nv)\, du\, dv\right) dx\, dy \\ & + \frac{p_V''^2}{2\pi^2} \sum_{\substack{m=1\\m>n}}^{+\infty} \sum_{\substack{n=1\\n>m}}^{+\infty} (-1)^{m+n} J_m J_n \iint_{x,y} B_D \exp\left[ik\left(\Delta + (x^2+y^2)/2d''\right)\right] \left(\iint_{u,v} \exp\left[ik\, p'_V(ux''+vy'')/2\sqrt{2}\pi d''\right] \cos(mu)\, du\, dv\right) \\ & \times \left(\iint_{u,v} \exp\left[ik\, p'_V(ux''+vy'')/2\sqrt{2}\pi d''\right] \cos(nv)\, du\, dv\right) dx\, dy \Bigg\} \end{aligned} \tag{A1}$$

Recognizing the integrals $\iint_{u,v} [-]du\, dv$ as Fourier transforms of cosine functions, it yields:

$$\begin{aligned} A''(x'',y'') = \Bigg\{ & J_0^2 \iint_{x,y} B_D \exp\left[ik\left(\Delta + (x^2+y^2)/2d''\right)\right] \delta(u,v)\, dx\, dy \\ & + J_0 \sum_{m=1}^{+\infty} (-1)^m J_m \iint_{x,y} B_D \exp\left[ik\left(\Delta + (x^2+y^2)/2d''\right)\right] \left[\delta(u-mp'',v) + \delta(u+mp'',v)\right] dx\, dy \\ & + J_0 \sum_{n=1}^{+\infty} (-1)^n J_n \iint_{x,y} B_D \exp\left[ik\left(\Delta + (x^2+y^2)/2d''\right)\right] \left[\delta(u,v-np'') + \delta(u,v+np'')\right] dx\, dy \\ & + \sum_{\substack{m=1\\m>n}}^{+\infty} \sum_{\substack{n=1\\n>m}}^{+\infty} (-1)^{m+n} J_m J_n \iint_{x,y} B_D \exp\left[ik\left(\Delta + (x^2+y^2)/2d''\right)\right] \left[\delta(u-mp'',v) + \delta(u+mp'',v)\right]\left[\delta(u,v-np'') + \delta(u,v+np'')\right] dx\, dy \Bigg\} \end{aligned} \tag{A2}$$

where $\delta(u,v)$ is the Dirac "delta" function and $p''$ is equal to: $p'' = \lambda f \sqrt{2}/p'_V$. Making use of the Dirac function properties, a new expression of $A''(x'',y'')$ comes as:

$$\begin{aligned} A''(x'',y'') = \Bigg\{ & J_0^2 B_D(x'',y'')\exp\left[ik\left(\Delta(x'',y'') + (x''^2+y''^2)/2d''\right)\right] \\ & + J_0 \sum_{m=1}^{+\infty} (-1)^m J_m B_D(x''-mp'',y'')\exp\left[ik\left(\Delta(x''-mp'',y'') + ((x''-mp'')^2+y''^2)/2d''\right)\right] \\ & + J_0 \sum_{m=1}^{+\infty} (-1)^m J_m B_D(x''+mp'',y'')\exp\left[ik\left(\Delta(x''+mp'',y'') + ((x''+mp'')^2+y''^2)/2d''\right)\right] \\ & + J_0 \sum_{n=1}^{+\infty} (-1)^n J_n B_D(x'',y''-np'')\exp\left[ik\left(\Delta(x'',y''-np'') + (x''^2+(y''-np'')^2)/2d''\right)\right] \\ & + J_0 \sum_{n=1}^{+\infty} (-1)^n J_n B_D(x'',y''+np'')\exp\left[ik\left(\Delta(x'',y''+np'') + (x''^2+(y''+np'')^2)/2d''\right)\right] \\ & + \sum_{\substack{m=1\\m>n}}^{+\infty} \sum_{\substack{n=1\\n>m}}^{+\infty} (-1)^{m+n} J_m J_n B_D(x''-mp'',y''-np'')\exp\left[ik\left(\Delta(x''-mp'',y''-np'') + ((x''-mp'')^2+(y''-np'')^2)/2d''\right)\right] \\ & + \sum_{\substack{m=1\\m>n}}^{+\infty} \sum_{\substack{n=1\\n>m}}^{+\infty} (-1)^{m+n} J_m J_n B_D(x''+mp'',y''-np'')\exp\left[ik\left(\Delta(x''+mp'',y''-np'') + ((x''+mp'')^2+(y''-np'')^2)/2d''\right)\right] \\ & + \sum_{\substack{m=1\\m>n}}^{+\infty} \sum_{\substack{n=1\\n>m}}^{+\infty} (-1)^{m+n} J_m J_n B_D(x''-mp'',y''+np'')\exp\left[ik\left(\Delta(x''-mp'',y''+np'') + ((x''-mp'')^2+(y''+np'')^2)/2d''\right)\right] \\ & + \sum_{\substack{m=1\\m>n}}^{+\infty} \sum_{\substack{n=1\\n>m}}^{+\infty} (-1)^{m+n} J_m J_n B_D(x''+mp'',y''+np'')\exp\left[ik\left(\Delta(x''+mp'',y''+np'') + ((x''+mp'')^2+(y''+np'')^2)/2d''\right)\right] \Bigg\} \end{aligned} \tag{A3}$$

Equation A3 also writes as:

$$\begin{aligned}
A''(x'',y'') = &\{J_0^2 B_D(x'',y'') \exp[ik(\Delta(x'',y'') + (x''^2 + y''^2)/2d'')] \\
&+ J_0 \exp[ik(x''^2 + y''^2)/2d''] \sum_{m=1}^{+\infty} (-1)^m J_m B_D(x'' - mp'', y'') \exp\left[ik\left(\Delta(x'' - mp'', y'') - \frac{mp''}{d''}x''\right)\right] \exp\left[ik\frac{m^2 p''^2}{2d''}\right] \\
&+ J_0 \exp[ik(x''^2 + y''^2)/2d''] \sum_{m=1}^{+\infty} (-1)^m J_m B_D(x'' + mp'', y'') \exp\left[ik\left(\Delta(x'' + mp'', y'') + \frac{mp''}{d''}x''\right)\right] \exp\left[ik\frac{m^2 p''^2}{2d''}\right] \\
&+ J_0 \exp[ik(x''^2 + y''^2)/2d''] \sum_{n=1}^{+\infty} (-1)^n J_n B_D(x'', y'' - np'') \exp\left[ik(\Delta(x'', y'' - np'')) - \frac{np''}{d''}y''\right] \exp\left[ik\frac{n^2 p''^2}{2d''}\right] \\
&+ J_0 \exp[ik(x''^2 + y''^2)/2d''] \sum_{n=1}^{+\infty} (-1)^n J_n B_D(x'', y'' + np'') \exp\left[ik\left(\Delta(x'', y'' + np'') + \frac{np''}{d''}y''\right)\right] \exp\left[ik\frac{n^2 p''^2}{2d''}\right] \\
&+ \exp[ik(x''^2 + y''^2)/2d''] \sum_{\substack{m=1 \\ m>n}}^{+\infty} \sum_{\substack{n=1 \\ n>m}}^{+\infty} (-1)^{m+n} J_m J_n B_D(x'' - mp'', y'' - np'') \exp\left[ik\left(\Delta(x'' - mp'', y'' - np'') + p''\frac{-mx'' - ny''}{d''}\right)\right] \exp\left[ik\frac{(m^2 + n^2)p''^2}{2d''}\right] \\
&+ \exp[ik(x''^2 + y''^2)/2d''] \sum_{\substack{m=1 \\ m>n}}^{+\infty} \sum_{\substack{n=1 \\ n>m}}^{+\infty} (-1)^{m+n} J_m J_n B_D(x'' + mp'', y'' - np'') \exp\left[ik\left(\Delta(x'' + mp'', y'' - np'') + p''\frac{mx'' - ny''}{d''}\right)\right] \exp\left[ik\frac{(m^2 + n^2)p''^2}{2d''}\right] \quad (A4) \\
&+ \exp[ik(x''^2 + y''^2)/2d''] \sum_{\substack{m=1 \\ m>n}}^{+\infty} \sum_{\substack{n=1 \\ n>m}}^{+\infty} (-1)^{m+n} J_m J_n B_D(x'' - mp'', y'' + np'') \exp\left[ik\left(\Delta(x'' - mp'', y'' + np'') + p''\frac{-mx'' + ny''}{d''}\right)\right] \exp\left[ik\frac{(m^2 + n^2)p''^2}{2d''}\right] \\
&+ \exp[ik(x''^2 + y''^2)/2d''] \sum_{\substack{m=1 \\ m>n}}^{+\infty} \sum_{\substack{n=1 \\ n>m}}^{+\infty} (-1)^{m+n} J_m J_n B_D(x'' + mp'', y'' + np'') \exp\left[ik\left(\Delta(x'' + mp'', y'' + np'') + p''\frac{mx'' + ny''}{d''}\right)\right] \exp\left[ik\frac{(m^2 + n^2)p''^2}{2d''}\right] \}
\end{aligned}$$

Assuming that the following approximations are made:

1. $p''$ is small with respect to the pupil diameter $D$, thus the transmission map of the pupil function $B_D(x'' \pm mp'', y'' \pm np'')$ reduces to $B_D(x'', y'')$ that can be set to unity over the full entrance aperture.

2. The wavefront error terms of the complex exponentials in Equation A4 are developed as function of their partial derivatives:

$$\Delta(x'' + mp'', y'' + np'') \approx \Delta(x'', y'') + mp'' \frac{\partial \Delta(x'', y'')}{\partial x''} + np'' \frac{\partial \Delta(x'', y'')}{\partial y''} \quad (A5)$$

   whatever are the indices $m$ and $n$:

3. Common WFE terms $\exp[ik\Delta(x'', y'')]$ can be removed from Equation A4 since they should disappear when computing the image intensities $I''(x'', y'') = |A''(x'', y'')|^2$.

4. For similar reason, the quadratic phase terms $\exp[ik(x''^2 + y''^2)/2d'']$ and $\exp[ik(m^2 + n^2)p''^2/2d'']$ can be removed from Equation A4.

Then inserting the relation A5 into Equation A4 and using elementary trigonometric formulas Equation A4 is rewritten with as:

$$\begin{aligned}
A''(x'',y'') = &\{J_0^2 + 2J_0 \sum_{m=1}^{+\infty} (-1)^m J_m \cos\left[k\left(mp''\Delta'_X + \frac{mp''}{d''}x''\right)\right] \\
&+ 2J_0 \sum_{n=1}^{+\infty} (-1)^n J_n \cos\left[k(np''\Delta'_Y) + \frac{np''}{d''}y''\right] \\
&+ 2\sum_{\substack{m=1 \\ m>n}}^{+\infty} \sum_{\substack{n=1 \\ n>m}}^{+\infty} (-1)^{m+n} J_m J_n \cos\left[ik\left(mp''\Delta'_X + np''\Delta'_Y + p''\frac{mx'' + ny''}{d''}\right)\right] \\
&+ 2\sum_{\substack{m=1 \\ m>n}}^{+\infty} \sum_{\substack{n=1 \\ n>m}}^{+\infty} (-1)^{m+n} J_m J_n \cos\left[ik\left(mp''\Delta'_X - np''\Delta'_Y + p''\frac{mx'' - ny''}{d''}\right)\right]
\end{aligned} \quad (A6)$$

which demonstrates the relation 8 in the main text.

# APPENDIX 2. EXPRESSIONS OF OFF-AXIS INTENSITIES

$$I''_1(x'',y'') = A+B-A\cos\left[kp''\left(\Delta'_{X'}+\Delta'_Y+\frac{x''+y''+2p''}{d''}\right)\right]+B\cos\left[kp''\left(\Delta'_{X'}-\Delta'_Y+\frac{x''-y''}{d''}\right)\right]+C\sin\left[2kp''\left(\Delta'_{X'}+\frac{x''+p''}{d''}\right)\right]+C\sin\left[2kp''\left(\Delta'_Y+\frac{y''+p''}{d''}\right)\right]$$

$$I''_2(x'',y'') = A+B+A\sin\left[kp''\left(\Delta'_{X'}+\Delta'_Y+\frac{x''+y''+2p''}{d''}\right)\right]-B\sin\left[kp''\left(\Delta'_{X'}-\Delta'_Y+\frac{x''-y''}{d''}\right)\right]+C\cos\left[2kp''\left(\Delta'_{X'}+\frac{x''+p''}{d''}\right)\right]+C\sin\left[2kp''\left(\Delta'_Y+\frac{y''+p''}{d''}\right)\right]$$

$$I''_3(x'',y'') = A+B+A\cos\left[kp''\left(\Delta'_{X'}+\Delta'_Y+\frac{x''+y''+2p''}{d''}\right)\right]-B\cos\left[kp''\left(\Delta'_{X'}-\Delta'_Y+\frac{x''-y''}{d''}\right)\right]-C\sin\left[2kp''\left(\Delta'_{X'}+\frac{x''+p''}{d''}\right)\right]+C\sin\left[2kp''\left(\Delta'_Y+\frac{y''+p''}{d''}\right)\right]$$

$$I''_4(x'',y'') = A+B+A\sin\left[kp''\left(\Delta'_{X'}+\Delta'_Y+\frac{x''+y''+2p''}{d''}\right)\right]+B\sin\left[kp''\left(\Delta'_{X'}-\Delta'_Y+\frac{x''-y''}{d''}\right)\right]+C\sin\left[2kp''\left(\Delta'_{X'}+\frac{x''+p''}{d''}\right)\right]+C\cos\left[2kp''\left(\Delta'_Y+\frac{y''+p''}{d''}\right)\right]$$

$$I''_5(x'',y'') = A+B+A\cos\left[kp''\left(\Delta'_{X'}+\Delta'_Y+\frac{x''+y''+2p''}{d''}\right)\right]+B\cos\left[kp''\left(\Delta'_{X'}-\Delta'_Y+\frac{x''-y''}{d''}\right)\right]+C\cos\left[2kp''\left(\Delta'_{X'}+\frac{x''+p''}{d''}\right)\right]+C\cos\left[2kp''\left(\Delta'_Y+\frac{y''+p''}{d''}\right)\right] \quad (A7)$$

$$I''_6(x'',y'') = A+B-A\sin\left[kp''\left(\Delta'_{X'}+\Delta'_Y+\frac{x''+y''+2p''}{d''}\right)\right]-B\sin\left[kp''\left(\Delta'_{X'}-\Delta'_Y+\frac{x''-y''}{d''}\right)\right]-C\sin\left[2kp''\left(\Delta'_{X'}+\frac{x''+p''}{d''}\right)\right]+C\cos\left[2kp''\left(\Delta'_Y+\frac{y''+p''}{d''}\right)\right]$$

$$I''_7(x'',y'') = A+B+A\cos\left[kp''\left(\Delta'_{X'}+\Delta'_Y+\frac{x''+y''+2p''}{d''}\right)\right]-B\cos\left[kp''\left(\Delta'_{X'}-\Delta'_Y+\frac{x''-y''}{d''}\right)\right]+C\sin\left[2kp''\left(\Delta'_{X'}+\frac{x''+p''}{d''}\right)\right]-C\sin\left[2kp''\left(\Delta'_Y+\frac{y''+p''}{d''}\right)\right]$$

$$I''_8(x'',y'') = A+B-A\sin\left[kp''\left(\Delta'_{X'}+\Delta'_Y+\frac{x''+y''+2p''}{d''}\right)\right]+B\sin\left[kp''\left(\Delta'_{X'}-\Delta'_Y+\frac{x''-y''}{d''}\right)\right]+C\cos\left[2kp''\left(\Delta'_{X'}+\frac{x''+p''}{d''}\right)\right]-C\sin\left[2kp''\left(\Delta'_Y+\frac{y''+p''}{d''}\right)\right]$$

$$I''_9(x'',y'') = A+B-A\cos\left[kp''\left(\Delta'_{X'}+\Delta'_Y+\frac{x''+y''+2p''}{d''}\right)\right]+B\cos\left[kp''\left(\Delta'_{X'}-\Delta'_Y+\frac{x''-y''}{d''}\right)\right]-C\sin\left[2kp''\left(\Delta'_{X'}+\frac{x''+p''}{d''}\right)\right]-C\sin\left[2kp''\left(\Delta'_Y+\frac{y''+p''}{d''}\right)\right]$$

# FIGURES

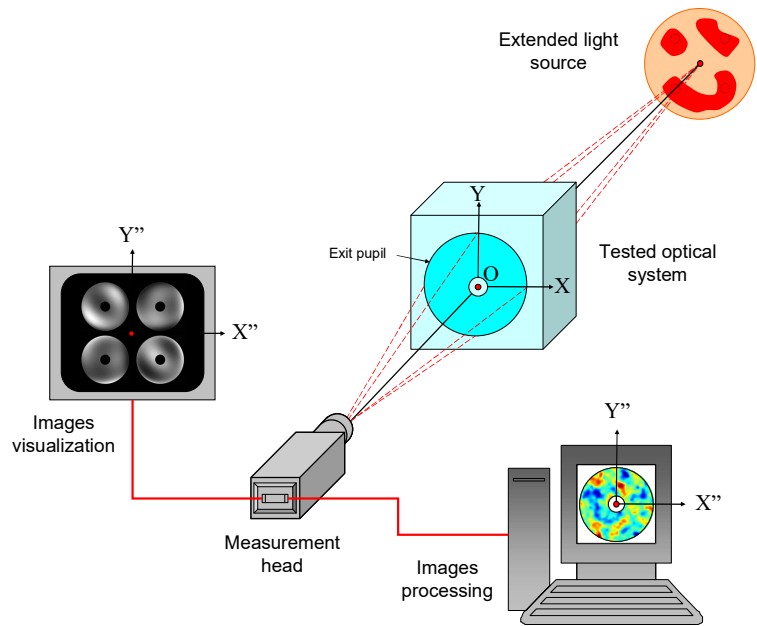

Figure 1: General concept of phase measurements using either GTF or GPF image filters [14].

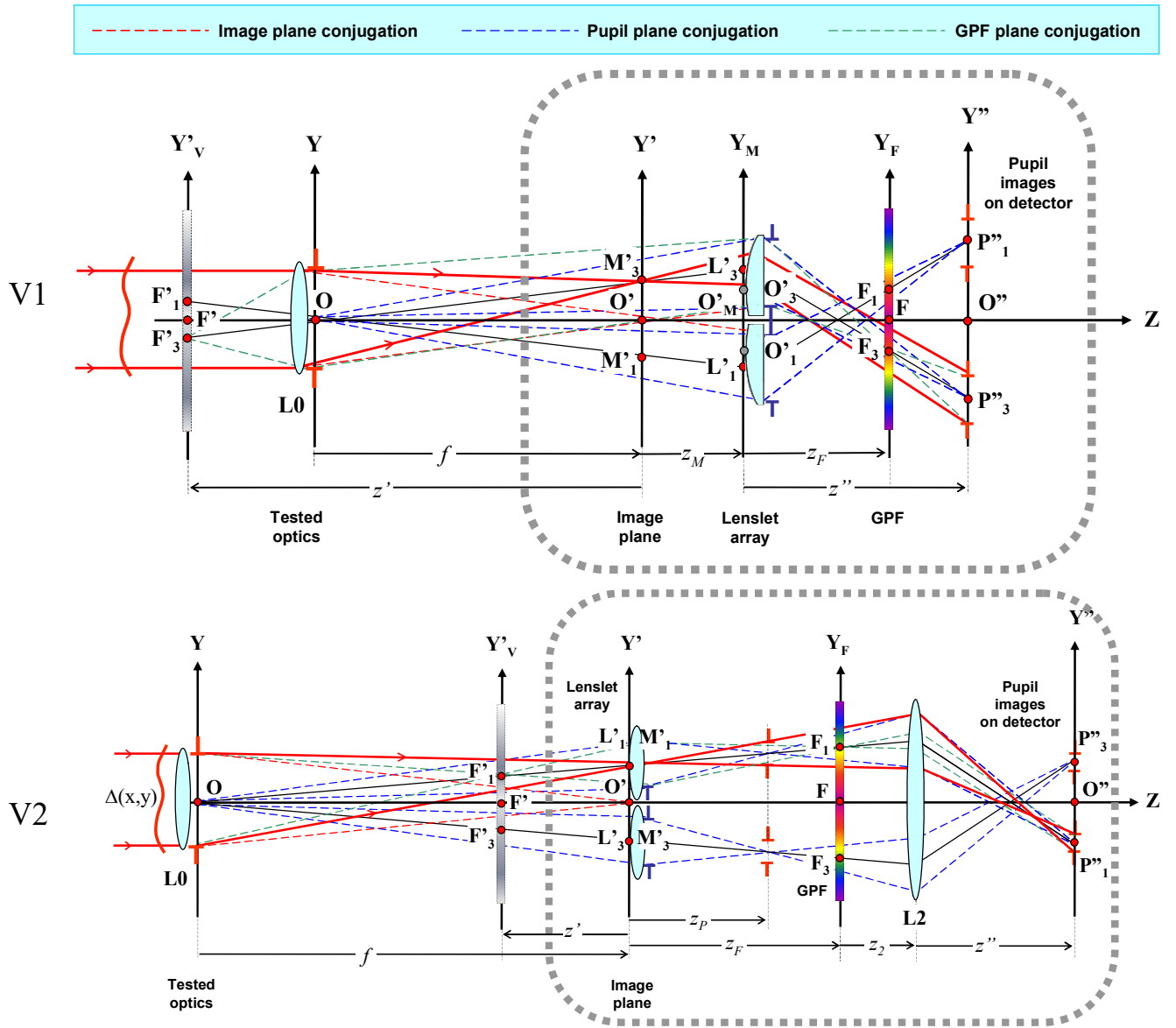

Figure 2: Two tentative optical designs of the CSPS optical measurement head.

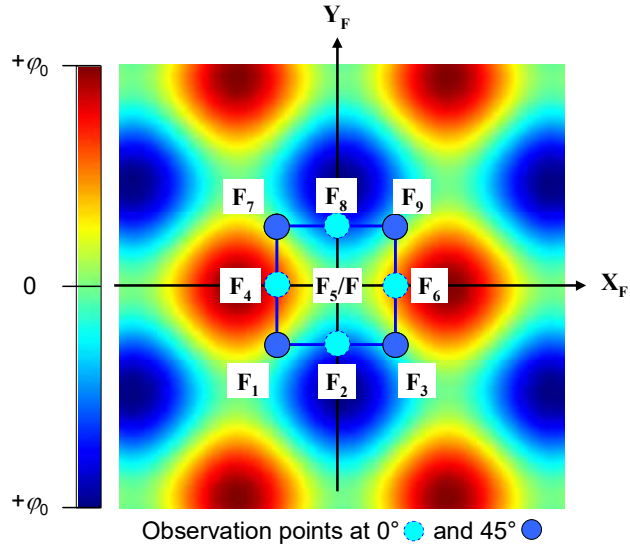

Figure 3: Phase map of the GPF in false colour, showing the locations of diffident observation points $F_i$ ($1 \leq i \leq 9$).

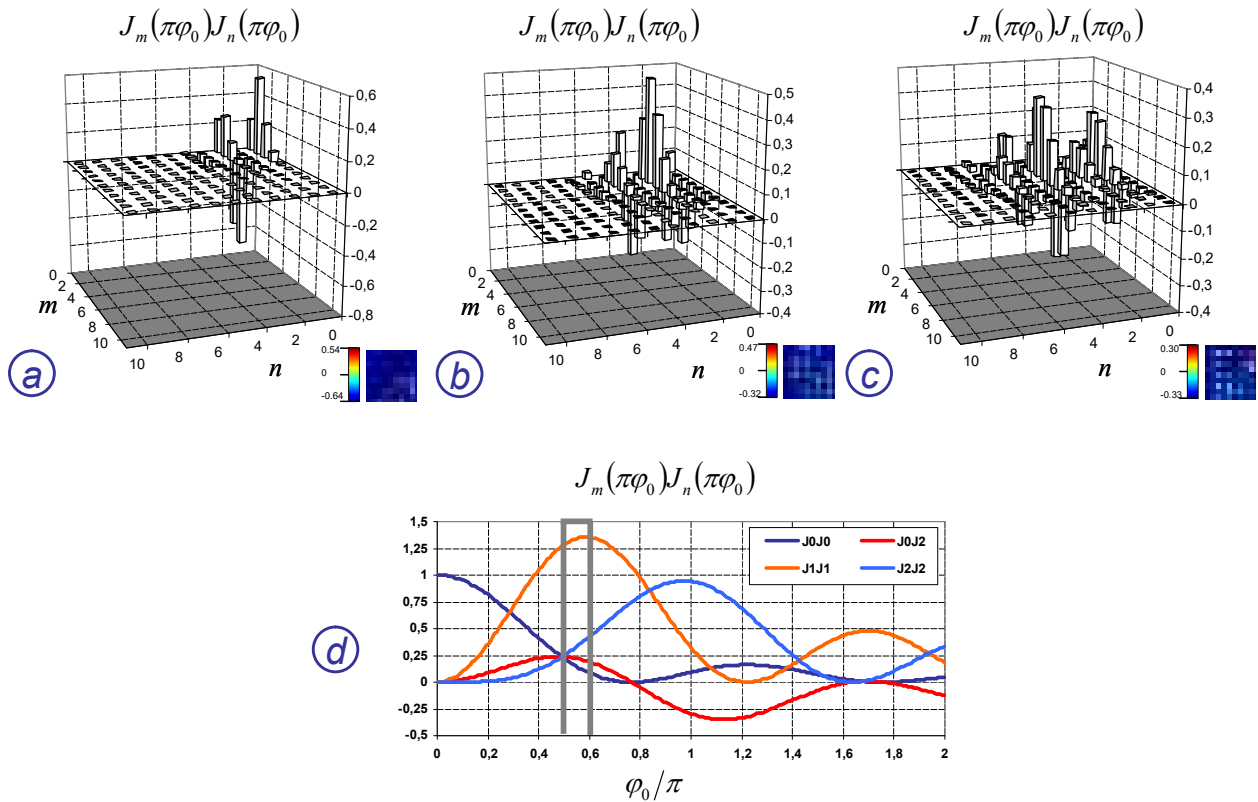

Figure 4: Graphical interpretation of Equation 10.
Histograms of the Bessel function products are illustrated for the cases when $\varphi_0 = \pi/2$ (a), $\varphi_0 = \pi$ (b) and $\varphi_0 = 3\pi/2$ (c). Figure 3-d shows their variations over the $[0, 2\pi]$ range. Grey bars in Figure 3-d indicate the best selected range.

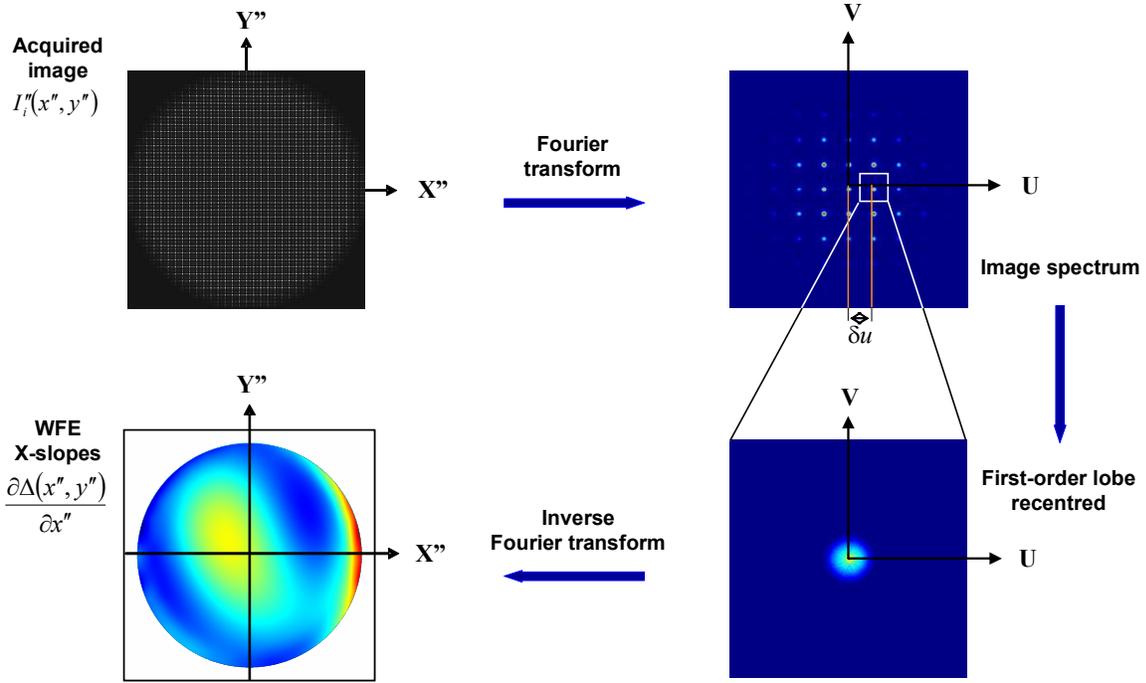

Figure 5: Double Fourier transform algorithm for reconstructing the slopes of the wavefront along X-axis.

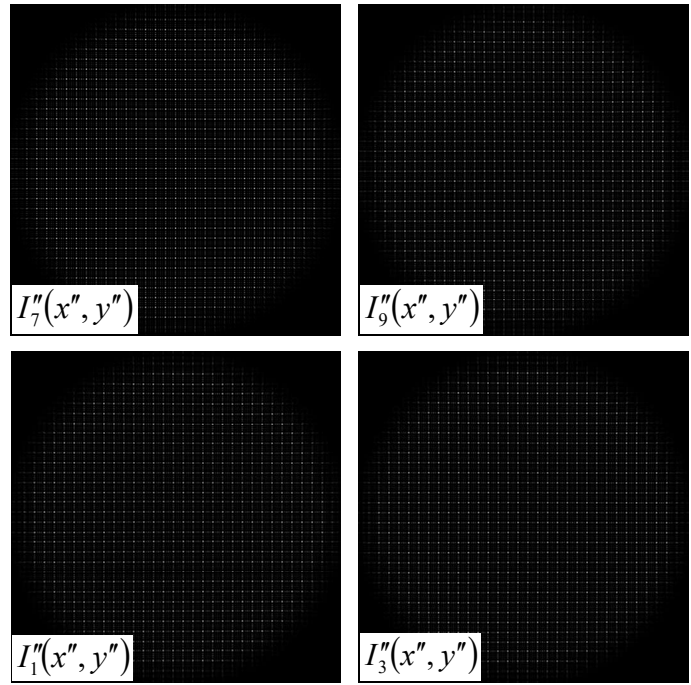

Figure 6: Simulation of four images acquired from the optical measurement head. Observation points are located at 45 degrees with respect to the X' and Y' axes. The dot effects in these figures result from the presence of the GPF.

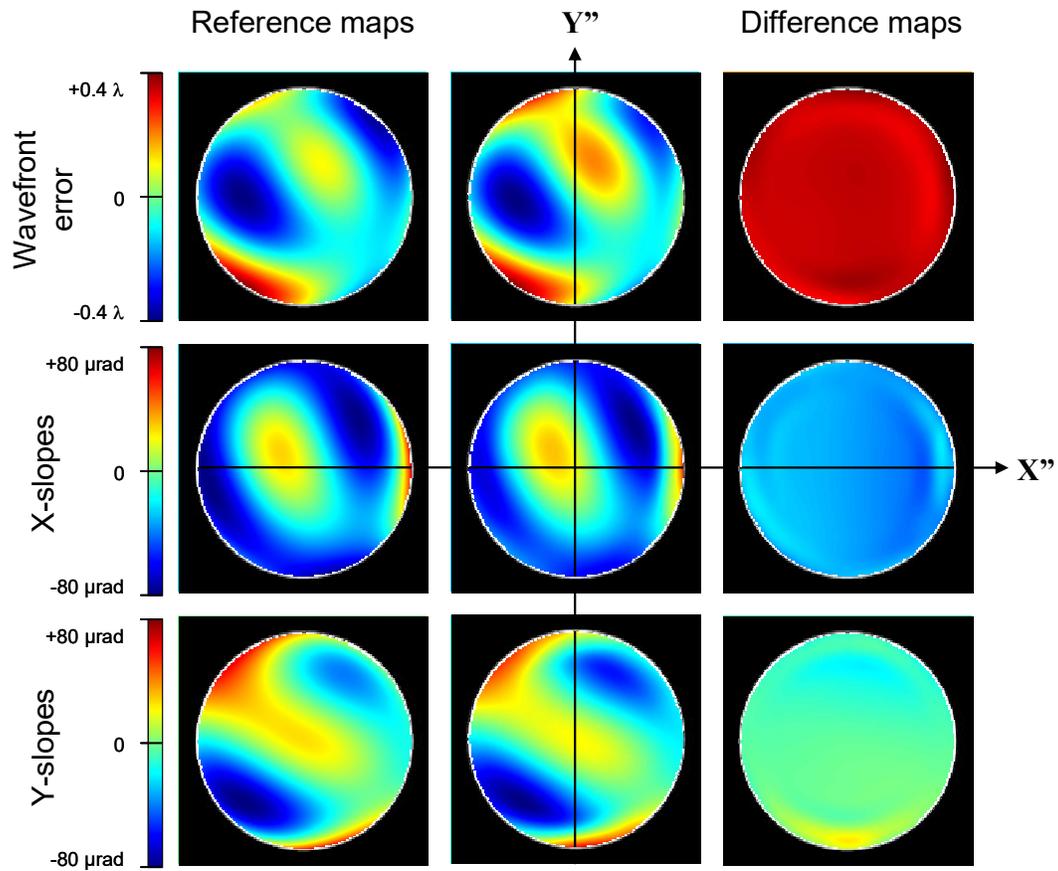

Figure 7: False colour simulations of the reference and measured wavefronts and of their slopes along the X" and Y" axes. Their difference maps are shown on the right column. Case of low order Zernike polynomials ≤ 16.

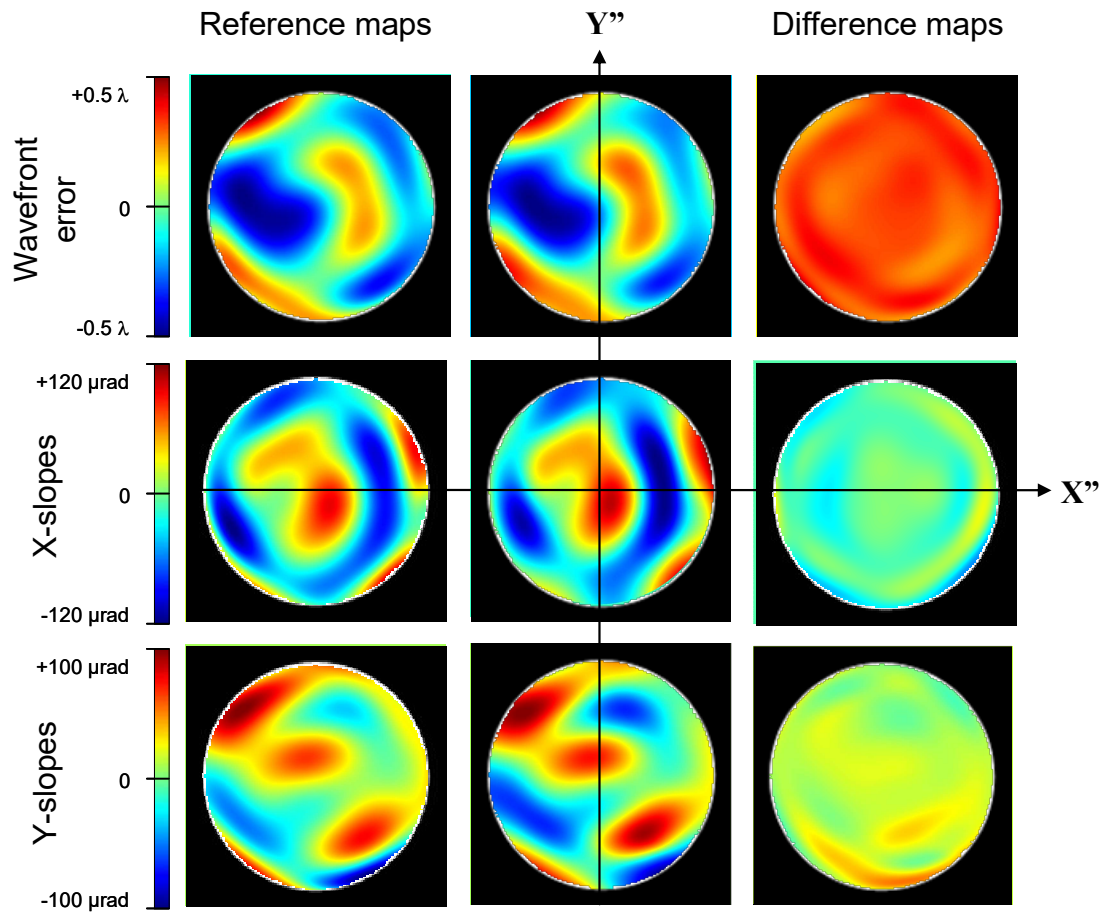

Figure 8: Same illustrations as in Figure 7. Case of mid-order Zernike polynomials ≤ 48.